\documentclass[conference,letterpaper]{IEEEtran}

\usepackage[utf8]{inputenc} 
\usepackage[T1]{fontenc}
\usepackage{url}
\usepackage{ifthen}
\usepackage{cite}
\usepackage[cmex10]{amsmath} 

\newcommand{\ONECOLUMN}[1]{}

\newcommand{\paperlink}[2]{}
\newcommand{\confpaperlink}[2]{}
\usepackage{setspace}
\usepackage{cite}
\usepackage[pdftex]{graphicx}
\usepackage{epstopdf}
\usepackage[cmex10]{amsmath}
\usepackage{amssymb}

\usepackage{arydshln} 

\usepackage{url}
\usepackage{epstopdf}
\usepackage{tikz}
\usetikzlibrary{automata,arrows,positioning,calc,%
shapes,chains}
\usepackage{mathtools}
\usepackage{texdef2015}
\newtheorem{theorem}{Theorem}

\newcommand{\Lcal}{\mathcal{L}}
\newcommand{\Qcal}{\mathcal{Q}}

\newcommand{\R}{\mathbb{R}}

\newcommand{\ql}{q_l}

\newcommand{\laml}[1][l]{\lambda^{(#1)}}
\newcommand{\muinv}{\mu^{-1}}

\newcommand{\smvec}[1]{\left[\begin{smallmatrix} #1\end{smallmatrix}\right]}
\newcommand{\rowvec}[1]{[\begin{matrix} #1\end{matrix}]}
\newcommand{\rvec}[1]{\smrvec{#1}}
\newcommand{\smrvec}[1]{\setlength\arraycolsep{2pt}\rowvec{#1}}

\newcommand{\piv}{\text{\boldmath{$\pi$}}}

\newcommand{\pibar}{\bar{\pi}}
\newcommand{\vvbar}{\bar{\vv}}

\newcommand{\onev}[1][n]{{\mathbf{1}}_{#1}}

\renewcommand{\vec}[1]{\begin{bmatrix} #1\end{bmatrix}}

\newcommand{\vvmbar}[2][m]{\bar{\vv}_{#2}}

\newcommand{\age}{\Delta}

\newlength{\swwidth}
\newcommand{\stw}[2]{\text{\settowidth{\swwidth}{\text{\ensuremath{#1}}}\makebox[\swwidth]{#2}}}
\newcommand{\vz}[1][0]{\stw{v_{00}}{#1}}
\newcommand{\xz}[1][0]{\stw{x_{0}}{#1}}

\newcommand{\Pc}{P_c}
\newcommand{\Pe}{P_e}
\newcommand{\slotage}{\age_{\text{slotted}}}

\title{Age of Information in Uncoordinated  Unslotted Updating}
\author{\IEEEauthorblockN{Roy~D.~Yates}
\IEEEauthorblockA{WINLAB, ECE Dept.,\\ Rutgers University\\
email: ryates@winlab.rutgers.edu}
\and
\IEEEauthorblockN{Sanjit~K.~Kaul}
\IEEEauthorblockA{Wireless Systems Lab,\\ IIIT-Delhi \\
email: skkaul@iiitd.ac.in}}
\IEEEoverridecommandlockouts

\begin{document}
\maketitle
\begin{abstract}
Sensor sources submit updates to a monitor  through an unslotted, uncoordinated, unreliable  multiple access collision channel. The channel is unreliable; a collision-free transmission is received successfully at the monitor with some transmission success probability. For an infinite-user  model in which the sensors collectively transmit updates as a Poisson process and each update has an independent exponential transmission time, 
a stochastic hybrid system (SHS) approach is used to derive the average age of information (AoI)  as a function of the offered load and the transmission success probability. The analysis is then extended to evaluate the individual age of a selected source. When the number of sources and update transmission rate grow large in fixed proportion, the limiting asymptotic individual age is shown to provide an accurate individual age approximation for a small number of sources. 
\end{abstract} 

\section{Introduction}
Consider a collection of sensors  that transmit updates to a central monitor. In many applications, complexity and energy considerations dictate that the sensors be transmit-only devices that blindly send update measurements without regard to the activity of other sensors \cite{Blaszczyszyn2008,Zhang2017FHMMF}. 
Because the transmit-only sources cannot coordinate, the transmissions are subject to collisions and the system operation is necessarily unslotted. 

 
Since the timeliness may be important, this work examines the age of information (AoI) of  these sensor updates. When the newest received update has time stamp $u(t)$, the age process is $\age(t)= t-u(t)$ \cite{Kaul-YG-Infocom2012} and the average age is $\limty{t} \E{\age(t)}$.

We note there has been growing interest in the AoI of sources sharing a communication facility, starting with multiple sources submitting updates through  queues \cite{Huang-Modiano-isit2015,Kadota-UBSM-Allerton2016,Kaul-Yates-isit2018priority,Yates-Kaul-IT2018, Najm-Telatar-aoi2018,maatouk2019age}. In addition, AoI has been analyzed for multiple users sharing a slotted system with various  levels of system coordination, including round-robin and Aloha-like contention \cite{Kaul-Yates-isit2017}, scheduled access \cite{Jiang-KZZN-isit2018,Hsu-isit2018,farazi2019fundamental,kosta2019age,maatouk2020optimality}, CSMA \cite{maatouk2019minimizing}, and random access with source-optimized contention  policies \cite{chen2019age}. However, age of information (AoI) in transmit-only sensor updates has not been studied. The graphical method of age analysis introduced in \cite{Kaul-YG-Infocom2012} and then employed  in e.g. \cite{Kam-KE-isit2013random,Kam-KE-isit2014diversity,Kam-KNE-IT2016diversity,Costa-CE-IT2016management,Costa-CE-isit2014,Champati-AG-aoi2018,Inoue-MTT-arxiv2017,Kam-KNWE-isit2016deadline,Feng-Yang-aoi2018} 
has not enabled age analysis of the collision channel.

\subsection{System Model}
In the collision channel, a transmission is collision-free if all other transmitters are idle during that transmission.  If an update suffers a collision, it is not received by the monitor. In addition, the communication channel is unreliable; a collision-free update will suffer an error and fail to be received by the monitor with probability $\Pe$.   

A key advantage of an unslotted system is that the transmission times can have arbitrary durations \cite{Gallager1985multiaccess}. To avoid a combinatorial explosion of the state space, we assume  the transmission times of the updates are modeled as independent exponential $(\mu)$ random variables.   Furthermore, the collection of sensors in aggregate  initiate update transmissions as a rate $\lambda$ Poisson point process. This is consistent with the ``infinite user'' model of historical importance in the analysis of the maximum stable throughput of collision resolution protocols \cite{abramson1970aloha,roberts1975aloha,gallager1978conflict, tsybakov1980random,Gallager1985multiaccess}.
\subsection{Paper Summary}

For the collection of uncoordinated sensors, we consider two types of age metrics. The {\em system age} is defined as the age of the most recent update received from any sensor in the system. For  the system age, an update from any sensor reduces the age at the monitor. This is in contrast to the {\em individual age} of a selected sensor  among  $N$ sensors.  Poisson arrivals of transmitted updates and exponential update  transmission times enable the method of stochastic hybrid systems (SHS) for age analysis. 
Section~\ref{sec:SHS}, provides a short introduction to the SHS method and then uses SHS  to analyze the system age in Section~\ref{sec:analysis}.  

Using the probability of correct detection $\Pc=1-P_e$, the system age analysis is extended to  evaluate the individual age in Section~\ref{sec:indiv}.  The individual age, in the limit of a large number of users and proportional  system service rate, is shown to converge to simple function of the offered load, that approximates the individual age even for a small number of sources.  The paper concludes with a discussion of open issues in Section~\ref{sec:conclusion}.
 
\section{Average System Age}
 \subsection{SHS Background}\label{sec:SHS}
 A  stochastic hybrid system (SHS) \cite{Hespanha-2006modelling} has state $[q(t),\xv(t)]$ such that $\xv(t)\in\R^{1\times n}$ and $q(t)\in\Qcal=\set{0,\ldots,M}$ is a continuous-time
 Markov chain. 

For AoI analysis, $q(t)$ describes the discrete state of a network while the age vector $\xv(t)$ describes the continuous-time evolution of a collection of  age-related processes.   
The SHS approach was introduced  in \cite{Yates-Kaul-IT2018}, where
it was shown  that age tracking can be implemented as a simplified SHS with non-negative linear reset maps in which the continuous state is a piecewise linear process 
\cite{Vermes-1980,Davis-1984,Deville-DDZ-siam2016moment}. For finite-state systems,  
this led to a set of age balance equations and simple conditions \cite[Theorem~4]{Yates-Kaul-IT2018}  under which $\E{\xv(t)}$ converges to a fixed point.  

A description of this simplified SHS for AoI analysis now follows.
In the graph representation of the Markov chain $q(t)$, each state $q\in\Qcal$ is a node and each transition $l$ is a directed edge $(q_l,q'_l)$ with transition rate $\laml$ from state $q_l$ to $q'_l$. 
Associated with each transition $l$, is  transition reset mapping  $\Amat_l\in\set{0,1}^{n\times n}$  that can induce a discontinuous jump $\xv'=\xv\Amat_l$ in the continuous state $\xv(t)$. 

Unlike an ordinary continuous-time Markov chain, the SHS Markov chain may include self-transitions in which the discrete state is unchanged because a reset occurs in the continuous state. Furthermore, for a given pair of states $i,j\in\Qcal$, there may be multiple transitions $l$ and $l'$ in which $q(t)$ jumps from $i$ to $j$ but the transition maps $\Amat_l$ and $\Amat_{l'}$ are different.

For each state $\qbar$, we denote the respective sets of incoming and outgoing  transitions by 
\begin{align}\eqnlabel{Lcalqbar}
\Lcal'_{\qbar}\!=\!\set{l\in\Lcal: q'_l=\qbar},\ \ 
\Lcal_{\qbar}\!=\!\set{l\in\Lcal: q_l=\qbar}.
\end{align}
Assuming the Markov chain $q(t)$ is ergodic, the discrete state Markov chain $q(t)$ has stationary probabilities 
 $\bar{\piv}=\rvec{\pibar_0&\cdots&\pibar_M}$ satisfying
\begin{align}
\bar{\pi}_{\qbar}\sum_{l\in\Lcal_{\qbar}}\laml&=\sum_{l\in\Lcal'_{\qbar}}\laml\bar{\pi}_{\ql},\quad \qbar\in\Qcal,
\eqnlabel{AOI-SHS-pi}
\end{align}
and the normalization constraint $\sum_{\qbar\in\Qcal}\bar{\pi}_\qbar=1$.
The next theorem provides a  way to derive the limiting average age vector $\E{\xv}=\limty{t}\E{\xv(t)}$.
\begin{theorem}\thmlabel{AOI-SHS}
\cite[Theorem~4]{Yates-Kaul-IT2018}
If the discrete-state Markov chain $q(t)$ is ergodic with stationary distribution $\bar{\piv}>0$ and there exists a non-negative vector $\vvbar=\rvec{\vvbar_0&\cdots\vvbar_M}$ 
such that 
\begin{align}
\bar{\vv}_{\qbar}\sum_{l\in\Lcal_{\qbar}}\laml &=\onev[]\bar{\pi}_{\qbar}+ \sum_{l\in\Lcal'_{\qbar}}\laml \bar{\vv}_{\ql}\Amat_l,\quad \qbar\in\Qcal,\eqnlabel{AOI-SHS-v}
\end{align}
then
the average age vector is 
$\E{\xv}=
\sum_{\qbar\in\Qcal} \vvbar_{\qbar}$.
\end{theorem}

In the next section,  \Thmref{AOI-SHS} is employed to find the average age
for uncoordinated unslotted updating.

\begin{figure}[t]
\centering
\begin{tikzpicture}[->, >=stealth', auto, semithick, node distance=2cm]
\tikzstyle{every state}=[fill=none,draw=black,thick,text=black,scale=0.9]
\node[state]    (0)                     {$0$};
\node[state] (1)[right of=0] {$1$};
\node[state] (2) [right of=1] {$2$};
\node  (H) [right of=2]   {$\cdots$};
\node[state] (M) [right of=H] {$M$};
\path
(0) 	edge[bend left=10,above]     node{\small $\lambda$}  (1)
(1) edge[bend left=10,below] node{\small $\Pc\mu$} (0)
(1) edge[bend left=60,below] node{\small $\Pe\mu$} (0)
(1) edge[bend left=10,above] node{\small $\lambda$} (2)
(2) edge[bend left=10,below] node{\small $2\mu$} (1)
(2) edge[bend left=10,above] node{\small $\lambda$} (H)
(H) edge[bend left=10,below] node{\small $3\mu$} (2)
(H) edge[bend left=10,above] node{\small $\lambda$} (M)
(M) edge[bend left=10,below] node{\small $M\mu$} (H);
\end{tikzpicture}
\caption{SHS Markov chain for the system age over an unslotted collision channel.}
\label{fig:unslotted-MC}
\vspace{-5mm}
\end{figure}
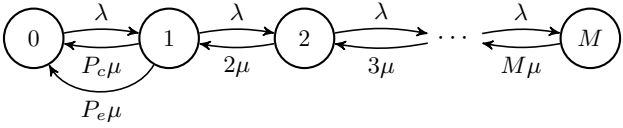

\subsection{SHS Analysis of the System Age}\label{sec:analysis}
For an SHS age model of the unslotted collision channel, the discrete state Markov chain for $q(t)$ is shown in Figure~\ref{fig:unslotted-MC} and the set of SHS transitions is given in Table~\ref{tab:overall-MC}. 
The discrete state $q(t)\in\set{0,1,2,\ldots}$ is the number of active transmitters. 
In the idle state $0$, the start of a transmission causes the system to jump to state $1$. This update is successfully delivered if it completes service before  another update begins transmission. Otherwise,
a jump to state $2$ begins a collision period in which  transmitted updates suffer collisions and are unsuccessful. In states $k\ge 2$, there are $k$ updates being transmitted in a $k$-way collision.   A collision period ends when the system returns to the idle state. 
\begin{table}[t]
\caption{SHS transitions for tracking the overall age in the Markov chain of Fig.~\ref{fig:unslotted-MC}.}\label{tab:overall-MC}
\begin{displaymath}\arraycolsep=2pt
\begin{array}{cccccc}
l & q_l\to q'_l & \laml &  \xv\Amat_l & \Amat_l & \vv_{\ql}\Amat_l\\ \hline
1 & 0\to 1 &\lambda	& \rvec{\xz&x_2}&\smvec{0 & 0\\0 & 1}&\rvec{\vz&v_{02}}\\
2 & 1\to 0 & \Pc\mu & \rvec{x_1&x_1}&\smvec{1 & 1\\ 0& 0} &\rvec{v_{11}&v_{11}}\\
3 & 1\to 0 & \Pe\mu & \rvec{x_2&x_2}&\smvec{0& 0\\1 & 1} &\rvec{v_{12}&v_{12}}\\
4 & 1\to 2 & \lambda & \rvec{x_2&x_2}&\smvec{0& 0\\1 & 1} &\rvec{v_{12}&v_{12}}\\
5 & 2\to 1 & 2\mu & \rvec{x_1&x_2}&\Imat&\rvec{v_{21}&v_{22}}\\
6 & 2\to 3 & \lambda & \rvec{x_1&x_2}&\Imat&\rvec{v_{21}&v_{22}}\\
7 & 3\to2 & 3\mu & \rvec{x_1&x_2}&\Imat&\rvec{v_{31}&v_{32}}\\
\vdots &\vdots&\vdots&\vdots&\vdots&\vdots\\
\vdots & M\!\to\! M\!-\!1 &M\mu	& \rvec{x_1&x_2}&\Imat&\rvec{v_{M1}&v_{M2}}
\end{array}
\end{displaymath}
\vspace{-5mm}
\end{table}

 The 
 age state is 
 $\xv(t)=\rvec{x_1(t)& x_2(t)}$ where $x_2(t)$ is the age at the monitor and $x_1(t)$ is what the age at the monitor would become if an update in service were to complete transmission at time $t$. Our objective is to calculate the average age at the monitor $\age=\limty{t}\E{x_2(t)}$. In each state $q(t) = \qbar$, the continuous state evolves according to 
$\dot{\xv}(t)=\onev[]=\rvec{1&1}$.



With respect to AoI, an age reduction in $x_2(t)$ occurs only  when a collision-free update is delivered successfully. 
In particular, in transition $l=1$, the system goes from idle to having a single update in service. In this case, the mapping $\xv'=\xv\Amat_1=\rvec{0 & x_2}$ resets $x_1$ to $x_1=0$, the age of the fresh update that just began transmission. On the other hand, $x'_2=x_2$ is unchanged because it tracks the age at the monitor. In state $1$, the transition $l=2$ corresponds to the update being transmitted collision-free  and also being successfully received. In this transition, $\xv'=\xv\Amat_2=\rvec{x_1 & x_1}$ resets $x_2$ to $x_2'=x_1$,
the age of the update that was just successfully received.  By contrast, transition $l=3$ corresponds to the update being transmitted collision-free but it fails to be received.  In this transition, $\xv'=\xv\Amat_3=\rvec{x_2 & x_2}$ leaves the age $x_2$ at the monitor unchanged. This transition also resets $x_1$ to $x_1'=x_2$, which destroys the ability of the update in transmission to reduce the age at the monitor. Similarly, $\Amat_4=\Amat_3$ because in transition $l=4$, a second update collides with an update in a transmission. Since this collision guarantees that neither update in transmission is successfully received,  this transition also sets $x_1'=x_2$,

While state $1$ has exactly one update being in service,  this update  may or may not be collision-free. This information is encoded in the continuous state $\xv(t)$. Specifically, $x_1(t)<x_2(t)$ when there is  single update with age $x_1(t)$ in the middle of a collision-free transmission at time $t$; otherwise $x_1(t)=x_2(t)$.
In particular, the transition $l=4$ into state $2$ initiates a collision period in which  $x_1(t)=x_2(t)$. 
This condition is preserved throughout the collision period, including when the system transitions through state $1$ and back to the idle state $0$. 
%

A consequence of the Poisson update arrival process  is that the number of updates being simultaneously transmitted can be arbitrarily large. However, in order to apply \Thmref{AOI-SHS}, the  state space is truncated so that the largest collision has $M$ updates.  The average age in the truncated system is $\age_M$. The average system age  with an infinite user population is $\age=\limty{M}\age_M$.


To employ \Thmref{AOI-SHS}, observe first that \eqnref{AOI-SHS-pi} implies
$\lambda\pibar_0=\mu\pibar_1$ and for $k=1,\ldots,M-1$,
\begin{align}
(\lambda+k\mu)\pibar_{k}&=\lambda\pibar_{k-1}+(k+1)\mu\pibar_{k+1}.
\end{align}
Solving for $\pibar_k$, $k=1,\ldots,M$, in terms of $\rho=\lambda/\mu$ and enforcing the normalization constraint yields
\begin{align}\eqnlabel{pibar-all}
\pibar_0 =\bigl(\sum_{j=0}^M \rho^j/j!\bigr)^{\!-1},\quad 
\pibar_{k}=\frac{\rho^k}{k!}\pibar_0.
\end{align}
From \eqnref{AOI-SHS-v}, we have for $\qbar\in\set{0,1,2,M}$ that
\begin{subequations}\eqnlabel{SHS-vector}
\begin{align}
\lambda\vvmbar{0}&=\onev[]\pibar_{0}+\Pc\mu\vvmbar{1}\Amat_2+\Pe\mu\vvmbar{1}\Amat_3,\\
(\lambda+\mu)\vvmbar{1}&=\onev[]\pibar_1+\lambda\vvmbar{0}\Amat_1+2\mu\vvmbar{2},\\
(\lambda+2\mu)\vvmbar{2}&=\onev[]\pibar_{2}+\lambda\vvmbar{1}\Amat_4+3\mu\vvmbar{3},\\
M\mu\vvmbar{M}&=\onev[]\pibar_{M}+\lambda\vvmbar{M-1},
\shortintertext{and for $\qbar=k\in\set{3,\ldots,M-1}$,} 
(\lambda+k\mu)\vvmbar{k}&=\onev[]\pibar_{k}+\lambda\vvmbar{k-1} +(k+1)\mu\vvmbar{k+1}.
\end{align}
\end{subequations}
\begin{figure}
\centering
\includegraphics{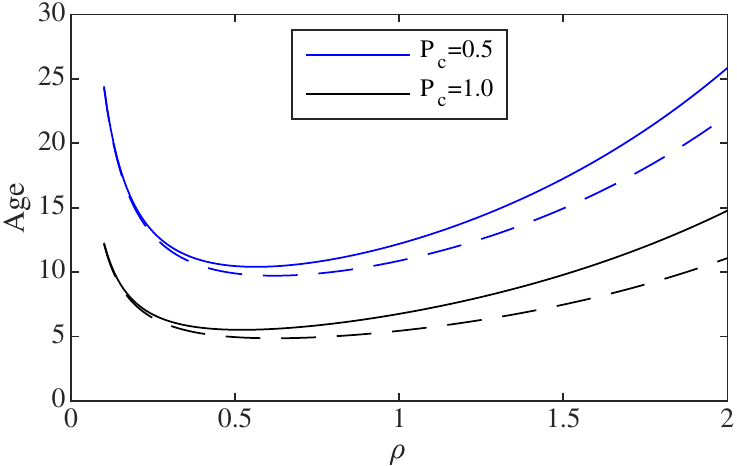}
\caption{Average system age $\age$ in \Thmref{collision-age} as a function of offered load $\rho$; time is normalized so that $\mu=1$. Dashed lines show the lower bound $\age_1(\rho)/(\mu P_c)$.}\label{fig:ageplot}
\vspace{-5mm}
\end{figure}
 Solving \eqnref{SHS-vector} for $\vvmbar{0},\ldots,\vvmbar{M}$, the average system age in the truncated system is \begin{align}
 \age_M&=\E{x_2}=\limty{M}\sum_{k=0}^M \vbar_{k2}.
 \end{align} 
 The average system age is then $\age=\limty{M}\age_M$. These steps can be found in the Appendix. For $j=1,2,\ldots$, we adopt the shorthand notation 
\begin{align}\eqnlabel{beta-gamma-defn}
\beta_j\equiv\sum_{i=j}^\infty\frac{\rho^i}{i!}e^{-\rho},\qquad  
\gamma_j
\equiv \sum_{k=0}^\infty 
\frac{j!}{(j+k)!}\rho^k, 
\end{align}
in order to state the following claim.\footnote{Note that $\beta_j=\prob{K\ge j}$ and $\gamma_j=\beta_j/\prob{K=j}$ for a Poisson $(\rho)$ random variable $K$,  While it is possible to state \Thmref{collision-age} with $\gamma_j$ replaced by $\beta_j/\prob{K=j}$, this  ratio of quantities that both go to zero as $j\to\infty$ can induce numerical stability issues in the calculation of $\age$.}
\begin{theorem}\thmlabel{collision-age}
Poisson updates through a collision channel achieve the average system age 
\begin{align*}
\age=\frac{(1\!+\!\rho)e^\rho}{\mu P_c \rho }\!+\!\frac{\beta_1}{\mu}
\!+\!\frac{(3\!+\!\rho)\beta_2}{2\mu}\!+\!\frac{\rho(1\!+\!\rho)\beta_2\gamma_3}{6\mu}
\!+\!\sum_{j=3}^\infty \frac{\beta_j\gamma_j}{j\mu}.
\end{align*}
\end{theorem}
\begin{figure}
\centering
\includegraphics{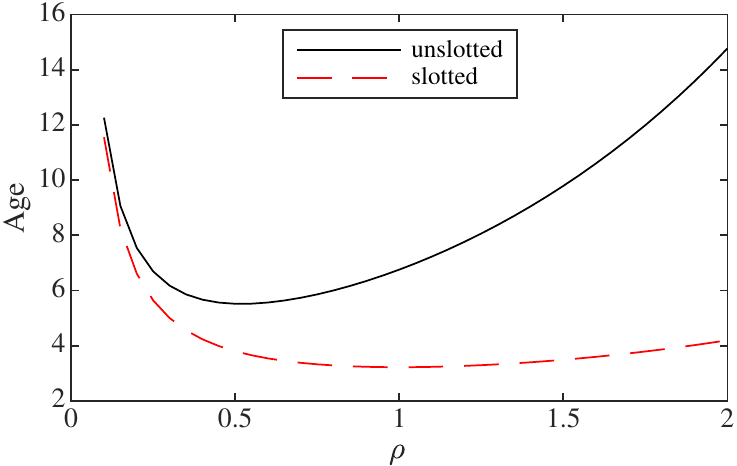}
\caption{Average system age $\age$ in \Thmref{collision-age} of the unslotted system 
vs.~the average system age $\slotage$ of the corresponding slotted Aloha system as a function of offered load $\rho$.}\label{fig:slotted}
\vspace{-5mm}
\end{figure}

With  time normalized so that $\mu=1$, Figure~\ref{fig:ageplot} depicts the system age in \Thmref{collision-age} as a function of the offered load $\rho$  for probability of correct reception $\Pc\in\set{0.5, 0.8,1}$. For all $\Pc$, the age becomes high when $\rho$  approaches zero or when $\rho$ becomes large and the system has too many collisions. For $\Pc=1$, the average  age happens to be minimized at $\rho=\rho^*=0.5195$, achieving the minimum age of $\age^*=5.513$. As $\Pc$ decreases, the optimal offered load increases slightly. For example, when $\Pc=0.5$, the optimal load is $\rho^*=0.5625$; this achieves an average age of $\age^*=10.40$. We see from Figure~\ref{fig:ageplot} that the average age is not particularly sensitive to variations in $\rho$ near $\rho^*$. 

We further observe that all terms in $\age$ are non-negative. With the definition
\begin{align}
\age_1(\rho)\equiv \paren{1+\frac{1}{\rho}}e^\rho, \eqnlabel{asymptotic-age}
\end{align}
the average system age satisfies the lower bound
\begin{align}
\age \ge \frac{\age_1(\rho)}{\mu P_c}
\end{align}
This simple lower bound, depicted in Figure~\ref{fig:ageplot} with  dashed lines, is tight for small $\rho$.

It is also instructive to compare the system age of the unslotted and slotted systems. Consider the corresponding infinite-user slotted system.  In each unit time slot, the number of fresh transmitted updates is a Poisson random variable $K$ with $\E{K}=\rho$. A fresh update is successfully transmitted in each time slot with probability $P_s=\prob{K=1}=\rho e^{-\rho}$. The average system age is \cite[Equation~(23)]{Kaul-Yates-isit2017}
\begin{align}
\slotage=\frac{1}{2}+\frac{1}{P_s}=\frac{1}{2}+\frac{e^\rho}{\rho}.
\end{align}
Figure~\ref{fig:slotted} compares average system age in the slotted and unslotted systems. We see that the age penalty for unslotted operation is negligible when the offered load $\rho$ is small. However, when the offered load is large, the age penalty becomes large because of the long collision periods induced by unslotted operation.

\section{Individual Age Analysis}\label{sec:indiv}
\begin{figure}[t]
\centering
\includegraphics{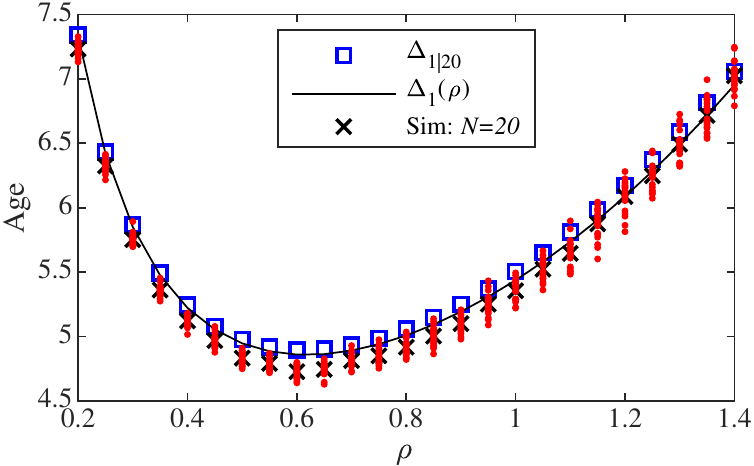}
\caption{Average individual age for $N=20$ sources. Time is normalized so that $\mu=20$. The figure compares $\age_{1|20}$ and $\age_1(\rho)$ against a simulation of $N=20$ on/off sources with aggregate update arrival rate $20\rho$. At each $\rho$, 
$\color{red}\bullet$ marks the time-average individual ages of each of the $20$ sources while $\times$ marks the average age averaged over the $20$ sources.}
\label{fig:naloha20}
\end{figure}
In practice, the number of sources $N$ will be finite and it is desirable to characterize the age process of an individual source.  Fortunately, the infinite user model of \Thmref{collision-age} can be employed to evaluate the individual age for one of $N$ sources by reinterpreting  $P_c$, the probability of correct detection of a collision-free update, as the probability that the collision-free update reduces the age of a selected user. Specifically, suppose the aggregate updating  rate $\lambda$ in the infinite user model is from $N$ independent sources, each offering updates as a  Poisson process of rate $\lambda/N$. In this case, a transmitted update belongs to a source $i$ with probability $1/N$. Hence, 
\Thmref{collision-age} can be employed with an update that is transmitted collision-free as belonging to source $i$ (and thus offering an age reduction for source $i$) with probability  $P_c=1/N$. This yields the individual age
\begin{align}\eqnlabel{indiv-age}
\age_{1|N}=\frac{N(1+\rho)e^\rho}{\mu\rho }&+\frac{\beta_1}{\mu}
+\frac{(3+\rho)\beta_2}{2\mu}\nn
&+\frac{\rho(1+\rho)\beta_2\gamma_3}{6\mu}
+\sum_{j=3}^\infty \frac{\beta_j\gamma_j}{j\mu}.
\end{align}

For fixed service rate $\mu$ and fixed offered load $\rho$, \eqnref{indiv-age} implies that the individual age grows linearly with the number of users $N$. This is not surprising since the system bandwidth, as embodied in the fixed service rate $\mu$, is shared among $N$ sources. However, to provide good age performance as $N$ becomes large, the system needs bandwidth to grow in proportion to $N$.  In this case, we assume the system has $N$ sources, each  offering updates at rate $\lambda_0$ but the system bandwidth grows with $N$ so that the service rate of a transmission is $\mu=N\mu_0$. The normalized offered load remains fixed at $\rho=(N\lambda_0)/(N\mu_0)=\lambda_0/\mu_0$.  A transmitted update belongs to the selected source with probability $P_c=1/N$.  We also assume time is normalized so that $\mu_0=1$. Under these conditions, we observe  as $N\to\infty$ that
\begin{align}
\age_{1|N}\to \age_1(\rho).
\end{align}
Here we can interpret $\age_1(\rho)$ as the individual age on a collision channel in the limit of  the number of sources becoming large and the transmission time of an update approaching zero. In this asymptotic limit, the individual average age is minimized at $\rho=(\sqrt{5}-1)/2=0.618$.\footnote{It can be shown that $\rho=0.618$ also maximizes the probability the system is transmitting a collision-free update.}

\begin{figure}[t]
\centering
\includegraphics{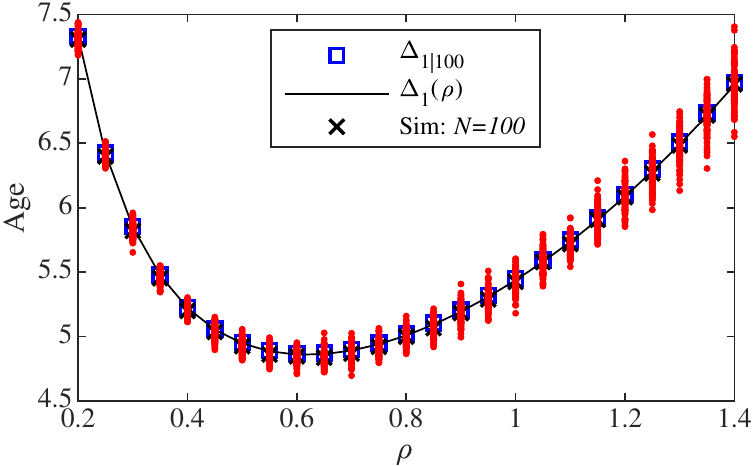}
\caption{Average individual age for $N=100$ sources. Time is normalized so that $\mu=100$. The figure compares $\age_{1|100}$ and $\age_1(\rho)$ against a simulation of $N=100$ on/off sources with aggregate update arrival rate $100\rho$. At each $\rho$, $\color{red}\bullet$ marks the time-average individual ages of each of the $100$ sources while $\times$ marks the average age averaged over the $100$ sources.}\label{fig:naloha100}
\end{figure}

 We will see this individual age model is somewhat pessimistic in that the Poisson update process of source $i$ can generate self-colliding updates; i.e., a source $i$ update can collide with time-overlapping updates also from source $i$. In practice, each source transmits one update at a time and never has a self-collision. In this sense, $\age_{1|N}$ 
 and $\age_1(\rho)$ 
 are approximations for the individual average age. 
 
 To evaluate these approximations, we simulate a system with $N$ independent on/off sources. Each source is either transmitting an update of exponential duration with expected value $1/\mu$, or being silent for an exponential period with expected length $1/\lambda_0-1/\mu$. By this construction, the two-state update process of each source offers updates at the longterm rate of $\lambda_0=\lambda/N$ updates per unit time.  
As $N$ becomes large, we expect the aggregate update process to be reasonably approximated by a rate $N\lambda_0=\lambda$ Poisson process. We also expect each source to obtain average individual age that is approximated by $\age_{1|N}$.

Under these conditions, Figures~\ref{fig:naloha20} and~\ref{fig:naloha100} compare $\age_{1|N}$ and $\age_1(\rho)$ against the simulated time-average ages experienced  by each of the $N$ on/off sources, each generating $50,000$ updates. Time is normalized so that $\mu=N$ and the average update transmission time is $1/\mu=1/N$. The aggregate offered load  is $\rho=(N\lambda_0)/N=\lambda_0$.

In Figure~\ref{fig:naloha20} with $N=20$ sources,  $\age_{1|N}$, which is derived from the infinite user model of \Thmref{collision-age},  is pessimistic in slightly (by 2-3\%) overestimating  the average age received by a source. The asymptotic approximation, $\age_1(\rho)$, which discards terms of $\age_{1|N}$ that become negligible as $\mu=N$ becomes large, is observed to be an even  better age approximation in the finite user system. In Figure~\ref{fig:naloha100} with $N=100$ sources, we see that that with more sources, the approximation  $\age_1(\rho)$ becomes an increasingly accurate  approximation to the average individual age. 

\section{Conclusion}\label{sec:conclusion}
For uncoordinated transmit-only sensors, this work provides an exact analysis for the system age. The uncoordinated transmit-only system works well as long as the normalized offered load is  near $\rho^*=0.6$. When these networks have  a nontrivial number of sources, $\age_1(\rho)$ is a useful approximation  for the individual age in a system with offered load $\rho$.  

From $\age_1(\rho)$, we see that the individual age penalty is substantial (on the order of $10\times$) if the offered load is, say,  $\rho^*/10$ or $10\rho^*$. Moreover, we saw in the comparison with the slotted system that age in the unslotted system is particularly sensitive to overloading the system. Configuring the network of transmit-only sources for the proper offered load would be an issue at time of deployment. On the other hand, adaptive configuration may also be possible if the  sources have access to some minimal feedback. 

In addition, there remain a number of open questions   about how additional coordination mechanisms,  such as collision detection and/or avoidance, and state-dependent updating policies, contribute to  reducing AoI. 




\appendix
\section*{Proof of \Thmref{collision-age}}
The mappings $\Amat_l$ induce $x_1(t)=x_2(t)$ in all states $k\neq 1$. This implies \eqnref{SHS-vector} has a solution such that $\vvmbar{k}=\rvec{\vbar_{k} &\vbar_{k}}=\vbar_{k}\onev[]$ for all $k\neq 1$. Only $\vvmbar{1}=\rvec{\vbar_{11} & \vbar_{12}}$ has distinct non-identical components. In terms of $\vbar_{11},\vbar_{12}$ and $\vbar_k$, $k\neq1$,  \eqnref{SHS-vector} becomes
\begin{subequations}\eqnlabel{SHS-scalar}
\begin{align}
\rho\vbar_{0}&=\muinv \pibar_{0}+\Pc\vbar_{11}+\Pe\vbar_{12},\eqnlabel{vbar0}\\
(1+\rho)\vbar_{11} &=\muinv\pibar_1+2\vbar_{2},\eqnlabel{vbar11}\\
(1+\rho)\vbar_{12}&=\muinv \pibar_1+\rho\vbar_0+2\vbar_{2},\eqnlabel{vbar12}\\
(2+\rho)\vbar_{2}&=\muinv \pibar_{2}+\rho\vbar_{12}\!+\!3\vbar_{3},\!
\eqnlabel{vbar2}\\
M\vbar_{M}&=\muinv \pibar_M+\rho\vbar_{M-1}.\eqnlabel{vbarM}
\shortintertext{and for  $3\le k\le M-1$,}
\rho_k\vbar_{k}&=\muinv \pibar_k + \rho\vbar_{k-1} +(k+1)\vbar_{k+1}\eqnlabel{vbark3plus}.
\end{align}
\end{subequations}
The average age   in \Thmref{AOI-SHS} becomes
\begin{align}\eqnlabel{ageM}
\age_M = \vbar_{0}+\vbar_{12} +\vbar_2+\sum_{j=3}^M\vbar_j.
\end{align}
In the limit of large $M$, we obtain the limiting average age $\age=\limty{M}\age_M$.
Equations \eqnref{vbarM} and \eqnref{vbark3plus}  admit the 
solution 
\begin{align}
\vbar_{k}=\frac{\rho}{k} \vbar_{k-1}+\frac{\beta_{k|M}}{k\mu}, \qquad 3 \le k\le M,\eqnlabel{vbark-step}
\end{align}
where 
\begin{align}\eqnlabel{betaM-defn}
\beta_{k|M}=\sum_{i=k}^M\pibar_i =\pibar_0\sum_{i=k}^M\rho^i/i!
\end{align}
is the stationary probability of the system being in a collision of $k$ or more updates. 
Now we observe that it follows from \eqnref{vbark-step} that for $3\le l\le M$,
\begin{align}\eqnlabel{vbarl}
\vbar_l =\sum_{j=3}^l \frac{(j-1)!}{\mu l!}\rho^{l-j}\beta_{j|M} +\frac{2\rho^{l-2}}{l!} \vbar_2.
\end{align}
Defining $V_{3:M}=\sum_{l=3}^M \vbar_l$, it then follows from \eqnref{vbarl} and reordering of the sums over $l$ and $j$ that
\begin{align}
\!\!\!V_{3:M}
&\!=\!\sum_{j=3}^M\!\frac{\beta_{j|M}}{j\mu}\sum_{l=j}^M\!\frac{j!}{l!}\rho^{l-j}\!+\!\frac{\rho\vbar_2}{3}\sum_{m=0}^{M-3} \frac{\rho^m}{(m\!+\!3)!}.
\eqnlabel{S3M-v2}
\end{align}
Defining 
$\gamma_{j|M}=\sum_{k=0}^{M-j}\frac{j!}{(k+j)!}\rho^k$,
the index shift $k=l-j$ in \eqnref{S3M-v2} yields
\begin{align}
V_{3:M}=\sum_{j=3}^M\frac{\beta_{j|M}}{j\mu}\gamma_{j|M}+\frac{\rho}{3}\gamma_{3|M}\vbar_2.\eqnlabel{S3M-v3}
\end{align}

Applying \eqnref{vbark-step} with $k=3$ to \eqnref{vbar2}  yields
\begin{align}
\vbar_2&= \frac{\pibar_{2}+\beta_{3|M}}{2\mu}+\frac{\rho}{2}\vbar_{12}=\frac{\beta_{2|M}}{2\mu}+\frac{\rho}{2}\vbar_{12}\eqnlabel{vbar2-v2}.
\end{align}
From  \eqnref{vbar2-v2} and the identity $\pibar_1+\beta_{2|M}=\beta_{1|M}$, it follows from \eqnref{vbar11} and \eqnref{vbar12} that
\begin{align}
\vbar_{11}&=\frac{\rho^2\vbar_0}{1+\rho}+\frac{\beta_{1|M}}{\mu},\quad
\vbar_{12}=\rho\vbar_0+\frac{\beta_{1|M}}{\mu}.\eqnlabel{vbar1-v2}
\end{align}
From \eqnref{vbar1-v2} and the identity $\Pc+\Pe=1$, it follows from \eqnref{vbar0} that 
\begin{align}
\vbar_0=\frac{1+\rho}{\mu\rho\Pc}.
\eqnlabel{vbar0-v2}
\end{align}
It then follows from \eqnref{vbar2-v2} and \eqnref{vbar1-v2}  that
\begin{align}\eqnlabel{vbar2-v3}
\vbar_2&=\frac{\rho^2}{2}\vbar_0 +\frac{\beta_{2|M}+\rho\beta_{1|M}}{2\mu}.
\end{align}
Applying \eqnref{S3M-v3}, \eqnref{vbar1-v2}, and \eqnref{vbar2-v3}  to 
\eqnref{ageM} 
and observing that  $\limty{M}\beta_{j|M}=\beta_j$ and 
$\limty{M}\gamma_{j|M}=\gamma_j$, the claim follows.
\newpage



\bibliographystyle{unsrt}
\bibliography{AOI-2019-01,collision}
\end{document}